\documentclass[onecolumn]{aa}
\usepackage{graphicx}
\usepackage{txfonts}
\def\ea{{\em et al.}}

\newcommand{\m}{\medbreak}
\newcommand{\EQ}{\begin{equation}}
\newcommand{\eq}{\end{equation}}

\newcommand{\eg}{\;=\;}

\begin{document}

   \title{Prospects for Dark Energy Evolution: \\ a Frequentist
Multi-Probe Approach }


   \author{ Ch. Y\`eche \inst{1}, A. Ealet\inst{2},
 A. R\'efr\'egier\inst{1}, C. Tao\inst{2}, A. Tilquin\inst{2},
                J.-M. Virey\inst{3}, \and D. Yvon \inst{1}
          }

   \institute{DSM/DAPNIA, CEA/Saclay, F-91191, Gif-sur-Yvette, France,
         \and
              Centre de Physique des Particules de Marseille, CNRS/IN2P3-Luminy and Universit\'e de la M\'editerran\'ee,
              Case 907, F-13288 Marseille Cedex 9, France,
          \and
              Centre de Physique Th\'eorique\thanks{``Centre de Physique Th\'eorique'' is UMR 6207 - ``Unit\'e Mixte
        de Recherche'' of CNRS and of the Universities ``de Provence'',
        ``de la M\'editerran\'ee'' and ``du Sud Toulon-Var''- Laboratory
        affiliated to FRUMAM (FR 2291).}, CNRS-Luminy and Universit\'e de Provence, Case 907,
              F-13288 Marseille Cedex 9, France.}

\date{\today}

\abstract{A major quest in cosmology is the understanding of the
nature of dark energy. It is now well known that a combination of
cosmological probes is required to break the underlying
degeneracies on cosmological parameters. In this paper, we present
a method, based on a frequentist approach, to combine probes
without any prior constraints, taking full account of the
correlations in the parameters. As an application, a combination
of current SNIa and CMB data with an evolving dark energy
component is first compared to other analyses. We emphasise the
consequences of the implementation of the dark energy
perturbations on the result for a time varying equation of state.
The impact of future weak lensing surveys on the measurement of
dark energy evolution is then studied in combination with future
measurements of the cosmic microwave background and type Ia
supernovae. We present the combined results for future mid-term
and long-term surveys and confirm that the combination with weak
lensing is very powerful in breaking parameter degeneracies. A
second generation of experiment is however required to achieve a
0.1 error on the parameters describing the evolution of dark
energy.

\keywords{cosmology: cosmological parameters -- supernovae -- CMB -- gravitational lensing -- large-scale structure in the universe -- dark energy -- equation of
state -- evolution} }
\authorrunning{Ch. Y\`eche \ea}
\titlerunning{Prospects for Dark Energy Evolution}
\maketitle

%

\section{Introduction}
\hspace{5mm}
Supernovae type Ia (SNIa) observations (Knop et al. \cite{SCP04},
Riess et al. \cite{Riess04}) provide strong evidence that the universe
is accelerating, in very good agreement with the WMAP Cosmic Microwave
Background (CMB) results (Bennett et al.~\cite{wmap}, Spergel et
al.~\cite{WMAPSpergel}) combined with measurements of large scale
structures (Hawkins et al. \cite{2dF}, Tegmark et
al. \cite{SDSSWMAP}).  The simplest way to explain the present
acceleration is to introduce a cosmological constant in Einstein's
equations. Combined with the presence of Cold Dark Matter, it forms
the so-called $\Lambda$CDM model. Even if this solution agrees well
with current data, the measured value of the cosmological constant is
very small compared to particle physics expectations of vacuum energy,
requiring a difficult fine tuning.  A favourite solution to this
problem involves the introduction of a new component, called "dark
energy" (DE), which can be a scalar field as in quintessence models
(Wetterich \cite{Wetterich}, Peebles \& Ratra \cite{PeeblesRatra}).

The most common way to study this component is to measure its
"equation of state" (EOS) parameter, defined as $w = p/\rho\,$, where
$p\,$ is the pressure and $\rho\,$ the energy density of the dark
energy. Most models predict an evolving equation $w(z)$.  It has been
shown (e.g., Maor et al.~\cite{Maor}, Maor et al.~\cite{Maor02},
Virey et al.~\cite{biasSN}, Gerke \&
Efstathiou~\cite{Gerke}) that neglecting such evolution biases the
discrimination between $\Lambda$CDM and other models.  The analysis of
dark energy properties needs to take time evolution (or redshift $z$
dependence) into account.

Other attractive solutions to the cosmological constant problem imply
a modification of gravity (for a review, cf., e.g., Lue et
al. \cite{Lue}, or Carroll et al. \cite{Carroll} and references
therein).  In this case, there is no dark energy as such and thus no
dark energy equation of state.  In this paper, we consider only the
dark energy solution, keeping in mind that Lue et al. (\cite{Lue}),
among others, have shown that the induced changes in the Friedmann
equations could be parameterised in ways very similar to a dark energy
evolving solution. \\

As various authors have noted (e.g., Huterer \& Turner
\cite{HutererTurner}, Weller \& Albrecht \cite{WA}), SNIa
observations alone will not be able to distinguish between an
evolving equation of state and $\Lambda$CDM.  This technique
indeed requires prior knowledge of the values of some parameters.
In particular, the precision on the prior matter density
$\Omega_m$ has an impact on the constraints on the time evolution
of the equation of state $w$, even in the simplest flat Universe
cosmology (e.g., Virey et al.~\cite{convergent}).

Extracting dark energy properties thus requires a combined
analysis of complementary data sets.  This can be done by
combining SNIa data with other probes such as the CMB, the large
scale distributions of galaxies, Lyman $\alpha$ forest data, and,
in the near future, the observation of large scale structure with
the Sunyaev-Zeldovich effect  (SZ) (Sunyaev \&
Zeldovich~\cite{SZ80}) or with weak gravitational lensing surveys
(WL), which provide an unique method to directly map the
distribution of dark matter in the universe (for reviews, cf.,
e.g., Bartelmann \& Schneider \cite{bar99}, Mellier et al.
\cite{mel02}; Hoekstra et al.~\cite{hoe02}, Refregier
\cite{alexandre}, Heymans et al. \cite{Heymans} and references
therein).

Many combinations have already been performed with different types
of data and procedures, (e.g., Bridle et al.
\cite{Bridle}, Wang \& Tegmark \cite{WangTegmark},
Tegmark et al.~\cite{SDSSWMAP}, Upadhye et al.~\cite{Upadhye},
Ishak \cite{Ishak}, Seljak et al.~\cite{Seljak},
Corasaniti et al.~\cite{Corasaniti}, Xia et al. \cite{ZhangXM}).  All
studies have shown the consistency of existing data sets with the
$\Lambda$CDM model and the complementarity of the different data sets
in breaking degeneracies and constraining dark energy for future
experiments. But the results differ by as much as 2$\sigma$ on the
central values of the parameters describing an evolving equation of
state.\\

In this paper, we have chosen three probes, which seem to best
constrain the parameters of an evolving equation of state when
combined, namely, SNIa, CMB and weak lensing. Considering a flat
Universe, we combine the data in a coherent way, that is to say,
under identical assumptions for the dark energy properties for the
three probes, and we completely avoid the use of priors. This had
not always been done systematically in all previous combinations.
We also adopt a frequentist approach for the data combination,
where the full correlations between the cosmological parameters
are taken into account. This method allows us to provide,
simultaneously, confidence intervals on a large number of distinct
cosmological parameters.  Moreover, this approach is very flexible
as it is easy to add or remove parameters in
contrast with other methods.\\

The paper is organised as follows: In Sec.~\ref{sec:Method}, we
describe our framework and statistical procedure, based on a
frequentist approach, which can accommodate all parameters without
marginalisation. For our simulation and analysis, we use the CMBEASY
package for CMB (Doran
\cite{CMBEASY}), the Kosmoshow program for SNIa  (Tilquin
\cite{kosmoshow}) and an extension of the calculations from Refregier et al.
(2003) for weak lensing. In each case, the programs take into account
the time evolution of the equation of state (cf
Sec.~\ref{sec:parameters} for details).

In Sec.~\ref{sec:FullFit}, we apply this method to current data
sets of SNIa and WMAP data.  We first verify that the constraints
on the cosmological parameters estimated with a Fisher matrix
technique (Fisher \cite{Fisher}), are consistent with those
obtained with a complete error analysis.  We then compare these
errors with other works and discuss the differences. In
particular, we discuss how the treatment of the dark energy
perturbations can explain some of the differences found in the
literature.

In Sec.~\ref{sec:MidTerm}, we study the statistical
sensitivities of different combinations of future surveys. We
simulate expectations for the ground surveys from the Canadian French
Hawaii Telescope Legacy Surveys (CFHTLS) and new CMB data from Olimpo
as well as the longer term Planck and SNAP space missions.
For these future experiments, the results are combined with a Fisher matrix
technique, compared and discussed.

Finally, our conclusions are summarised in Sec.~\ref{sec:conclusion}.

\section{Combination method}
\label{sec:Method}
\hspace{5mm}
In this section, we first summarise the framework used in this paper, and
describe our approach based on frequentist statistics.

\subsection{Dark Energy Parametrization}
\label{sec:DEParam}
\hspace{5mm}
The evolution of the expansion parameter is given by the Hubble parameter $H$ through the
Friedmann equation

\EQ
  \left(\frac{H(z)}{H_0}\right) ^2 \eg (1+z)^3\,\Omega_{m}+
    {\rho_X(z)\over \rho_X(0)} \, \Omega_{X}+(1+z)^2\,\Omega_{k}, \\
\eq

with

\EQ \label{rho} {\rho_X(z)\over \rho_X(0)} \eg \exp \left[
3\int_0^z\,\left(1+w(z')\right)\, d\,\ln (1+z') \right] \eq \m
where the ratio of the dark energy density to the critical density
is denoted $\Omega_X$ in a general model and  $\Omega_{\Lambda}$
in the simplest case of a Cosmological Constant ($w = -1$).
$\Omega_M$ is the corresponding parameter for (baryonic+cold dark)
matter. Note that we have neglected the radiation component
$\Omega_R$. The present total and curvature density parameters are
$\Omega$ and $\Omega_{\kappa}=1-\Omega$, respectively. The present
value of the Hubble constant is parameterised as $H_{0}=100 h$ km
s$^{-1}$ Mpc$^{-1}$.

As it is not possible to constrain a completely unknown functional
form $w(z)$ of the time evolution of the equation of state, we
adopt a parametric representation of the $z$ dependence of the
equation of state. We need this parametric form to fit all the
data sets over a large range of $z$: from $z\simeq 0-1$ for the
SNIa and weak lensing, up to $z\simeq 1100$ for the CMB. For this
purpose, we choose the parametrization proposed by Chevallier \&
Polarski  (\cite{Polarskiparam}) and Linder (\cite{LinderA}) :
\begin{center}
\EQ
w(z) = w_0 + {w_a}z/(1+z),
\label{eq:w0wa}
\eq
\end{center}
which has an adequate asymptotic behaviour. In this paper, we thus use
two parameters, $w_0$ and $w_a$, to describe the time evolution of the
equation of state (see justifications in Linder \& Huterer
~\cite{Linder05}).  For this parametrization of $w(z)$, Eq.~\ref{rho}
reduces to:
\begin{center}
\EQ
\rho_X(z)\eg \rho_X(0) \, e^{-3w_az/(1+z)} \, (1+z)^{3(1+w_0+w_a)}.
\eq
\end{center}
For a constant $w\equiv w_0$ ($w_a=0$),
the usual form $\rho_X(z)\eg \rho_X(0) \ (1+z)^{3(1+w_0)} $ is recovered.

The comoving distance $\chi$ is defined as
\begin{center}
\EQ
 \chi(z)= \int_0^z\,\frac{c}{H(z')}dz',
\eq
\end{center}
and the comoving angular-diameter distance $r(\chi)$  is  equal, respectively, to $\chi$,
$R_0 sin(\chi/R_0)$, $R_0\sinh(\chi/R_0)$, for a flat, closed and open
Universe where the present curvature radius of the universe is defined as
$R_0= c/(\kappa H_0)$ with respectively $\kappa^2 \equiv 1$, $-\Omega_{\kappa}$, and
$\Omega_{\kappa}$.

\subsection{Statistical approach}
\label{sec:parameters}
\hspace{5mm}
Most recent CMB analysis use Markov Chains
Monte Carlo simulations (Gilks et al.~\cite{MCMC}, Christensen \&
Meyer~\cite{MCMC_CMB}) with bayesian inference. The philosophical
debate between the bayesian and the frequentist statistical approaches
is beyond the scope of this paper (for a comparison of the two
approaches see, for instance, Feldman \& Cousins~\cite{Frequentist}
and Zech~\cite{Zech}). Here, we briefly review the principles of each
approach.

For a given data set, the bayesian approach computes the
probability distribution function (PDF) of the parameters
describing the cosmological model.  The bayesian probability is a
measure of the plausibility of an event, given incomplete
knowledge. In a second step, the bayesian constructs a 'credible'
interval, centered near the sample mean, tempered by 'prior'
assumptions concerning the mean.  On the other hand, the
frequentist determines the probability distribution of the data as
a function of the cosmological parameters and gives a confidence
level that the given interval contains the parameter. In this way,
the frequentist completely avoids the concept of a PDF defined for
each parameter. As the questions asked by the two approaches are
different, we might expect different confidence intervals.
However, the philosophical difference between the two methods
should not generally lead, in the end, to major differences in the
determination of physical parameters and their confidence
intervals when the parameters stay in a physical region.

Our work is based on the 'frequentist' (or 'classical') confidence
level method originally defined by Neyman (\cite{Neyman}).  This
choice avoids any potential bias due to the choice of priors.  In
addition, we have also found ways to improve the calculation
speed, which gives our program some advantages over other bayesian
programs. Among earlier combination studies (e.g., Bridle et al.
\cite{Bridle}, Wang \& Tegmark \cite{WangTegmark}, Tegmark et
al.~\cite{SDSSWMAP}, Upadhye et al.~\cite{Upadhye}, Ishak
\cite{Ishak}, Seljak et al.~\cite{Seljak}, Corasaniti et
al.~\cite{Corasaniti}, Xia et al. \cite{ZhangXM}) only that of
Upadhye et al. (\cite{Upadhye}) uses also a frequentist approach.

\subsubsection{Confidence levels with a frequentist approach}
\hspace{5mm}
For a given cosmological model defined by the $n$ cosmological
parameters $\theta=(\theta_{1},\ldots,\theta_{n})$, and for a data set
of $N$ quantities $x =(x_1,\ldots,x_N)$ measured with gaussian
experimental errors $\sigma_{x} =(\sigma_{1},\ldots,\sigma_N)$, the
likelihood function can be written as:
\begin{center}
\begin{equation}
{\cal L}(x,\sigma_{x};\theta) = \frac{1}{\sqrt{2\pi}\sigma_i}
exp\biggl(
-\frac{(x_i - x_{i,model})^2}{2\sigma_i^2}
\biggr).
\end{equation}
\end{center}
where $x_{model} =(x_{1,model},\ldots,x_{N,model})$
is a set of corresponding model dependent values.

In the rest of this paper, we adopt a $\chi^2$ notation, which means
that the following quantity is minimised:
\begin{center}
\begin{equation}
\chi^2(x,\sigma_{x};\theta) = -2 \ln ({\cal L}(x,\sigma_{x};\theta))
\end{equation}
\end{center}
We first determine the minimum $\chi^2_0$ of $\chi^2(x,\sigma_{x};\theta)$
letting free all the cosmological parameters.
Then, to set a confidence level (CL) on any individual
cosmological parameter $\theta_i$, we scan the variable $\theta_i$:
for each fixed value of $\theta_i$,
we minimise again $\chi^2(x,\sigma_{x};\theta)$ but with $n-1$ free parameters.
The $\chi^2$ difference, $\Delta \chi^2(\theta_i)$, between the new minimum
and  $\chi^2_0$, allows us to compute the CL on the variable,
assuming that the experimental errors are gaussian,
 \begin{center}
 \begin{equation}
1-{\rm CL}(\theta_i) = \frac{1}{\sqrt{2^{N_{dof}}} \Gamma(N_{dof}/2)}
\int_{\Delta \chi^2(\theta_i)}^{\infty}e^{-t/2}t^{N_{dof}/2 -  1}dt
\label{Eq:CL}
\end{equation}
\end{center}
where $\Gamma$ is the gamma function and the number of degrees of freedom $N_{dof}$
is equal to 1.
This method can be easily extended to two variables. In this case, the minimisations are
performed for $n-2$ free parameters and the confidence level ${\rm CL}(\theta_i,\theta_j)$ is
derived from Eq.~\ref{Eq:CL} with $N_{dof}=2$.

By definition, this frequentist approach does not require any
marginalisation to determine the sensitivity on a single individual
cosmological parameter.  Moreover, in contrast with bayesian
treatment, no prior on the cosmological parameters is needed.  With
this approach, the correlations between the variables are naturally
taken into account and the minimisation fit can explore the whole phase space
of the cosmological parameters.

In this study, the minimisations of $\chi^2(x,\sigma_{x};\theta)$ are
performed with the MINUIT package (James \cite{minuit}).
For the 9 parameter study proposed in this paper, each fit
requires around 200 calculations of $\chi^2$. The consumed
CPU-time is dominated by the computation of the angular power
spectrum ($C_{\ell}$) of the CMB in CMBEASY (Doran
~\cite{CMBEASY}). In practice, to get the CL for one variable, as
shown, for instance, in Fig.~\ref{fig:CLw0wa}, the computation of
the $C_{\ell}$ is done around 10000 times.  The total number of
calls to perform the study presented in
Tab.~\ref{tab:DataResults}, is typically 3 or 4 times smaller than
the number of calls in the MCMC technique used by Tegmark et al.
(\cite{SDSSWMAP}).  This method is very powerful for studying the
impacts of the parameters:  it is not costly to add or remove
parameters because the number of  $C_{\ell}$ computations scales
with the number of parameters, in contrast with the MCMC method,
which requires the generation of a new chain.

\subsubsection{Combination of cosmological probes with Fisher matrices}
\hspace{5mm} In parallel with this frequentist approach, to study
the statistical sensitivities of different combinations of future
surveys, we perform a prospective analysis based on the Fisher
matrix technique (Fisher \cite{Fisher}). We validate this approach
by comparing its estimates of the statistical errors for the
current data set with those obtained with the frequentist method
described above.

The statistical errors on the $n$ cosmological parameters
$\theta=(\theta_{1},\ldots,\theta_{n})$ are determined by using
the inverse of the covariance matrix $V$ called the Fisher matrix
$F$ defined as:
\begin{center}
\begin{equation}
\label{eq:Fisher}
    (V^{-1})_{ij}= F_{ij} = -\frac{\partial^{2}\ln
    {\cal L}(x;\theta)}{\partial\theta_{i}\partial\theta_{j}},
\end{equation}
\end{center}
where ${\cal L}(x;\theta)$ is the likelihood function depending on the
$n$ cosmological parameters and a data set of $N$ measured quantities
$x =(x_1,\ldots,x_N)$. A lower bound, and often a good estimate, for
the statistical error on the cosmological parameter $\theta_i$ is
given by $(V_{ii})^{1/2}$.

When the measurements of several cosmological probes are combined, the
total Fisher matrix $F_{tot}$ is the sum of the three Fisher matrices
$F_{SN}$, $F_{WL}$ and $F_{CMB}$ corresponding respectively to the
SNIa, weak lensing and CMB observations. The total covariance matrix
$F_{tot}^{-1}$ allows us to estimate both, the expected sensitivity on
the cosmological parameters, with the diagonal terms, and the
correlations between the parameters, with the off-diagonal terms.  The
Fisher matrices for each probe are computed as follows.

\paragraph*{\bf CMB:\, } In the case of CMB experiments, the data set
vector $x$ corresponds to the measurements of $C_{\ell}$, the
angular power spectrum of the CMB from $\ell=2$ to some cutoff
$\ell_{max}$. Using Eq.~\ref{eq:Fisher}, the Fisher matrix is
written as
\begin{center}
\begin{equation}
(F_{CMB})_{ij} = \sum_{l=2}^{\ell_{max}}
\frac{1}{\sigma_{C_{\ell}}^2} \cdot \frac{\partial
C_{\ell}}{\partial\theta_{i}}\cdot \frac{\partial
C_{\ell}}{\partial\theta_{j}}
\end{equation}
\end{center}
where $\sigma_{C_{\ell}}$ is the statistical error on $C_{\ell}$
obtained directly from published results or estimated as (see
Knox~\cite{knox95}):
\begin{center}
\begin{equation}
\label{eq:cmberror}  \sigma_{C_{\ell}} =
\sqrt{\frac{2}{(2\ell+1)f_{sky}}}\biggl[ C_{\ell} +
(\theta_{fwhm}s)^2\cdot
e^{\frac{\ell^2 \theta_{fwhm}^2}{8\ln(2)}} \biggr]
\end{equation}
\end{center}
where the second term incorporates the effects of instrumental noise
and beam smearing. In Eq.~\ref{eq:cmberror}, $\theta_{fwhm}$,   $f_{sky}$, and $s$ are
respectively the angular resolution, the fraction of the sky observed
and the expected sensitivity per pixel.

The $C_{\ell}$ and their derivatives with
respect to the various cosmological parameters are computed with
CMBEASY (Doran ~\cite{CMBEASY}), an object oriented C++ package
derived from CMBFAST (Seljak \& Zaldarriaga ~\cite{cmbfast}).

\paragraph*{\bf SNIa:\, }
\label{sneparams} The SNIa apparent magnitudes $m$ can be
expressed as a function of the luminosity distance as
\begin{center}
\begin{equation}
m(z)= M_{s_{0}} + 5log_{10}(D_L)
\end{equation}
\end{center}
where $D_L(z) \equiv (H_0/c)\;d_L(z)$ is the \emph{$H_0$-independent}
luminosity distance to an object at redshift $z$. The usual luminosity
distance $d_L(z)$ is related to the comoving angular-diameter distance
$r(\chi)$ by $d_L(z)= (1+z) \cdot r(\chi)$, with the definition of
$r(\chi)$ and $\chi(z)$ given in Sec.~\ref{sec:DEParam}. The
normalisation parameter $M_{s_{0}}$ thus depends on $H_0$ and on the
absolute magnitude of SNIa.

The Fisher matrix, in this case, is related to the measured
apparent magnitude $m_k$ of each object and its statistical error
$\sigma_{m_k}$ by
\begin{center}
\begin{equation}
(F_{SN})_{ij}= \sum_{k} \frac{1}{\sigma_{m_k}^2} \cdot \frac{\partial
m_k}{\partial\theta_{i}}\cdot \frac{\partial
m_k}{\partial\theta_{j}}.
\end{equation}
\end{center}

\paragraph*{\bf Weak lensing:\, }

The weak lensing power spectrum is given by (e.g., Hu \&
Tegmark~\cite{HuTeg99}, cf, Refregier~\cite{alexandre} for
conventions)
\begin{center}
\begin{equation}
\label{eq:cl}
C_{\ell} = \frac{9}{16} \left( \frac{H_{0}}{c}
\right)^{4} \Omega_{m}^{2}
  \int_{0}^{\chi_h} d\chi~\left[ \frac{g(\chi)}{a r(\chi)} \right]^{2}
  P\left(\frac{\ell}{r}, \chi\right),
\end{equation}
\end{center}
where
$r(\chi)$ is the  comoving angular-diameter distance, and
$\chi_{h}$ corresponds to the comoving distance to horizon.
The radial weight function $g$ is given by
\begin{center}
\begin{equation}
g(\chi) = 2 \int_{\chi}^{\chi_{h}} d\chi'~n(\chi')
   \frac{r(\chi)r(\chi'-\chi)}{r(\chi')},
\end{equation}
\end{center}
where $n(\chi)$ is the probability of finding a galaxy at comoving
distance $\chi$ and is normalised as $\int d\chi~n(\chi) =1$.

The linear matter power spectrum $P(k,z)$ is computed using the
transfer function from Bardeen et al. (\cite{bar86}) with the
conventions of Peacock (\cite{pea97}), thus ignoring the
corrections on large scales for quintessence models (Ma et
al.~\cite{Ma99}).
The linear growth factor of the matter overdensities
$\delta$ is given by the well known equation:
\begin{center}
\begin{equation}
\ddot{\delta}+2H\dot{\delta}-\frac{3}{2}H^2\Omega_m(a)\delta=0,
\end{equation}
\end{center}
where dots correspond to time derivatives, and $\Omega_m(a)$ is
the matter density parameter at the epoch corresponding to the
dimensionless scale factor $a$. This equation is integrated
numerically with boundary conditions given by the matter-dominated
solution, $G=\delta/a=1$ and $\dot{G}=0$, as $a \rightarrow 0$
(see eg. Linder \& Jenkins \cite{LinderJenkins}). We enforce the
CMB normalisation of the power spectrum $P(k,0)$ at $z=0$ using
the relationship between the WMAP normalisation parameter $A$ and
$\sigma_8$ given by Hu~(\cite{Hu04}). Considerable uncertainties
remain for the non-linear corrections in quintessence models (cf.
discussion in Hu~(\cite{Hu02})).  Here, we use the fitting formula
from Peacock \& Dodds (\cite{pea96}).

For a measurement of the power spectrum, the Fisher matrix element is
defined as:
\begin{center}
\begin{equation}
(F_{WL})_{ij}=  \sum_{\ell} \frac{1}{\sigma_{C_{\ell}}^2}
 \frac{\partial C_{\ell}}{\partial \theta_{i}} \frac{\partial
C_{\ell}}{\partial \theta_{j}},
\end{equation}
\end{center}
where the summation is over modes $\ell$ which can be reliably
measured.  This expression assumes that the errors
$\sigma_{C_{\ell}}$ on the lensing power spectrum are gaussian and
that the different modes are uncorrelated. Mode-to-mode
correlations have been shown to increase the errors on
cosmological parameters (Cooray \& Hu ~\cite{coo01}) but are
neglected in this paper.

Neglecting non-gaussian corrections, the statistical error
$\sigma_{C_{\ell}}$ in measuring the lensing power spectrum
$C_{\ell}$ (cf., e.g., Kaiser~\cite{b13}, Hu \& Tegmark
\cite{HuTeg99}, Hu \cite{Hu02}) is given by:
\begin{center}
\begin{equation}
\sigma_{C_{\ell}} =\sqrt{ \frac{2}{(2l+1) f_{\rm
sky}}}\left(C_{l}+\frac{\sigma_{\gamma}^{2}}{2n_{g}} \right),
\label{eq:wlerror}
\end{equation}
\end{center}
where $f_{\rm sky}$ is the fraction of the sky covered by the survey,
$n_{g}$ is the surface density of usable galaxies, and
$\sigma_{\gamma}^{2} =\langle |\gamma|^{2} \rangle$ is the shear
variance per galaxy arising from intrinsic shapes and measurement
errors.

\subsection { Cosmological parameters and models}
\hspace{5mm}
For the studies presented in this paper, we limit ourselves to
the 9 cosmological parameters: $\theta = \Omega_b, \Omega_m , h, n_s,
\tau, w_0, w_a, A$ and $M_{s_0} $, with the following standard definitions:\\
- ($\Omega_i$ , i=b,m) are densities for baryon and matter respectively
($\Omega_{m}$ includes both dark matter and baryons),\\
- $h$ is the Hubble constant in units of 100 km/s/Mpc,\\
- $n_s$ is the spectral index of the primordial power spectrum,\\
- $\tau$ is the reionisation optical depth, \\
- $A$ is the normalisation parameter of the power spectrum for
CMB and weak lensing (cf  Hu \& Tegmark (\cite{HuTeg99}) for definitions).
 The matter power spectrum is normalised according to the COBE
normalisation (Bunn \& White \cite{BunnWhite}), which corresponds
to $\sigma_{8}=0.88$. This is consistent with the WMAP results
(Spergel et al. \cite{WMAPSpergel}) and with the average of recent
cosmic shear measurements (see compilation tables in Mellier et
al. \cite{mel02}, Hoekstra et al. \cite{hoe02}, Refregier \cite{alexandre}). \\
- $M_{s_{0}}$ is the normalisation parameter from SNIa
(cf Sec.~\ref{sneparams}), \\
- Dark energy  is described by the $w_0$ parameter corresponding to
the value of the equation of state at $z$=0.
When the $z$ dependence of the equation of state is studied,
an additional parameter $w_a$ is defined (cf Sec.~\ref{sec:DEParam}). \\

The reference fiducial model of our simulation is a $\Lambda$CDM
model with parameters $\Omega_{m}=0.27$, $\Omega_{b}=0.0463$,
$n_s=0.99$, $h=0.72$, $\tau= 0.066$, $A= 0.86$, consistent with
the WMAP experiment (see tables 1-2 in Spergel et al.
\cite{WMAPSpergel}).  In agreement with this experiment, we assume
throughout this paper that the universe is flat, i.e.,
$\Omega=\Omega_{m}+\Omega_{X}=1$.  We also neglect the effect of
neutrinos, using 3 degenerate families of neutrinos with masses
fixed to 0.

In the following, we will consider deviations from this reference
model.  For the equation of state, we use as a reference $w_0 = -0.95$
and $w_a = 0$ as central values (we have not used exactly $w_0=-1$ to
avoid transition problems in the CMB calculations). To estimate the
sensitivity on the parameters describing the equation of state, we
also consider two other fiducial models: a SUGRA model, with ($
w_0=-0.8, w_a=0.3$) as proposed by, e.g., Weller \& Albrecht
(\cite{WA}) to represent quintessence models, and a phantom model
(Caldwell~\cite{Caldwell}) with ($w_0 = -1.2, w_a=-0.3$).

In this analysis, the full covariance matrix on all parameters is used
with no prior constraints on the parameters,
avoiding biases from internal degeneracies.
We have implemented the time evolving parametrization of the equation of state
in simulations and analysis of the three probes we consider
in this paper, i.e. CMB, SNIa and weak lensing.

\section {Combination of current surveys}
\label{sec:FullFit} \hspace{5mm} We first apply our statistical
approach to the combination of recent SNIa and CMB data, without
any external constraints or priors.  The comparison of the
statistical errors obtained with a global fit using this
frequentist treatment, with those predicted with the Fisher matrix
technique, also allows us to validate the procedure described in
Sec.~\ref{sec:Method}. Finally, we compare our results with other
published results.

\subsection{Current surveys}
\hspace{5mm} We use the 'Gold sample' data compiled by Riess et
al. (\cite{Riess04}), with 157 SNIa including a few at $z > 1.3$
from the Hubble Space Telescope (HST GOODS ACS Treasury survey),
and  the published data from WMAP taken from Spergel et al.
(\cite{WMAPSpergel}).

We perform two distinct analyses: in the first case, the equation
of state is held constant with a single parameter $w_0$ and we fit
8 parameters, as described in Sec.~\ref{sec:parameters}; in the
second case, the $z$ dependence of the equation of state is
modelled by two variables  $w_0$ and $w_a$ as defined in
Sec.~\ref{sec:DEParam}, and we fit 9 parameters.

\subsection{Results}
\begin{figure}
   \centering
   \includegraphics[width=8cm]{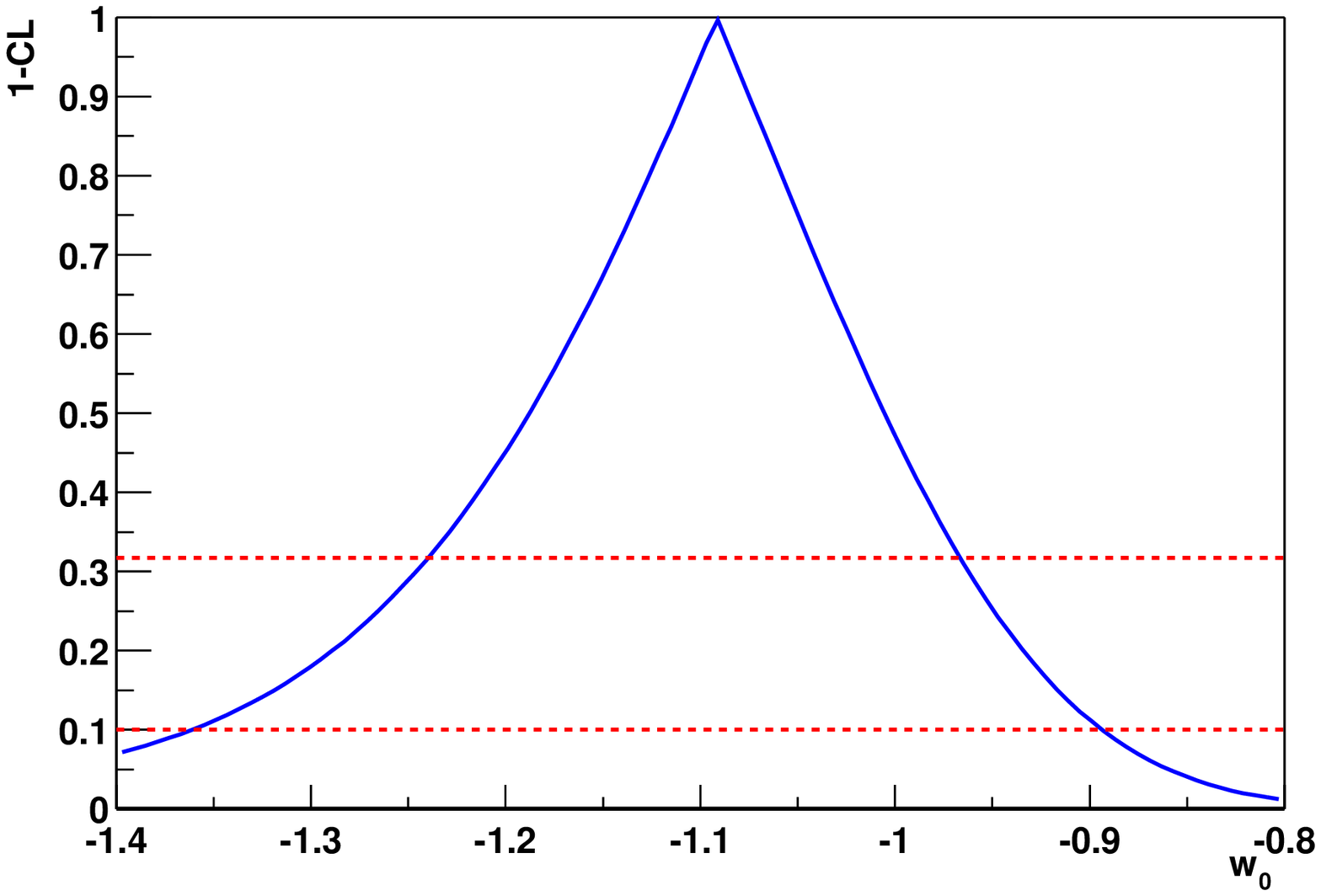}
   \includegraphics[width=8cm]{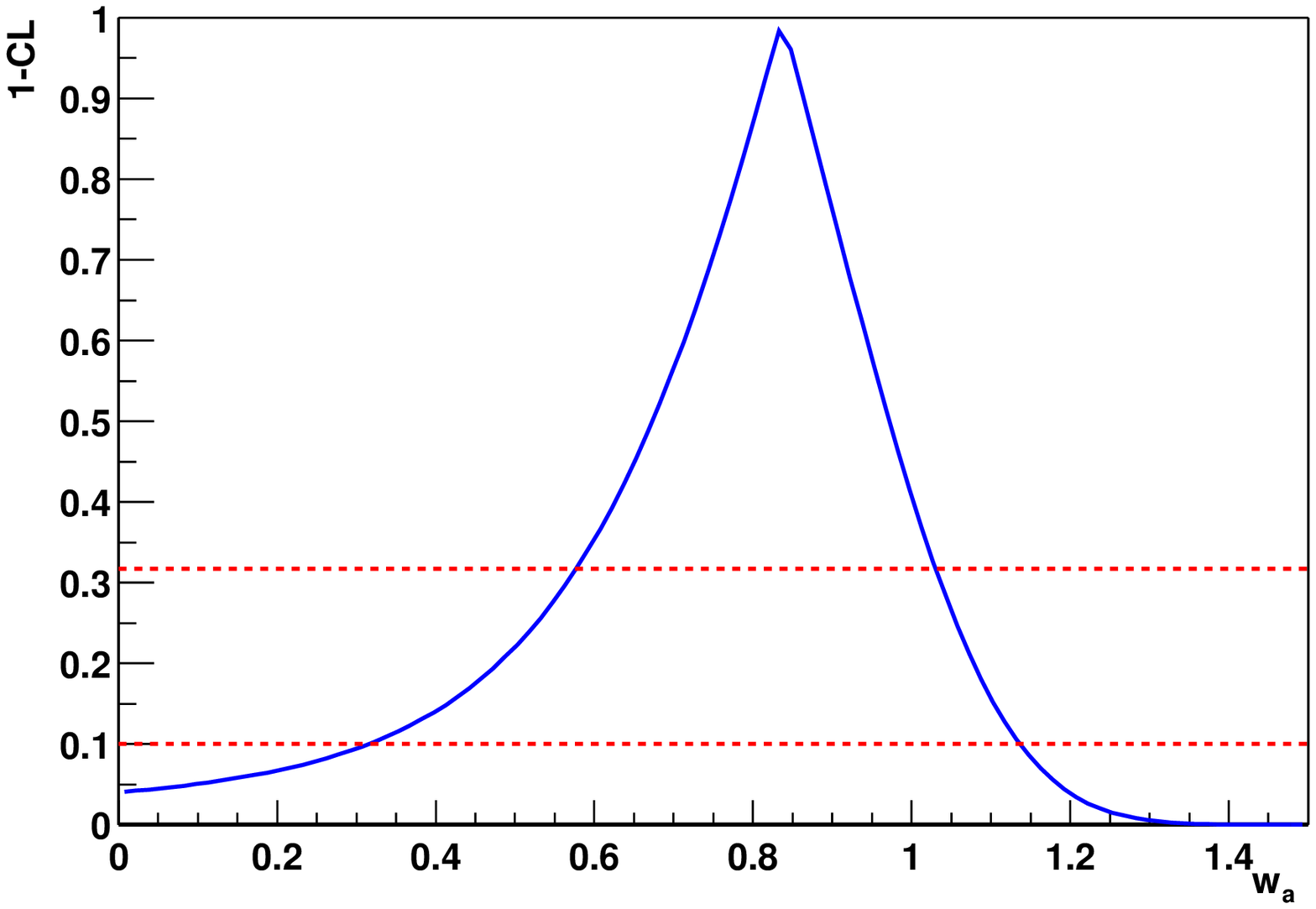}
   \caption{\footnotesize Confidence level (CL) plots on parameters $w_0$ (left)
   and $w_a$ (right),
   using WMAP and Riess et al.~\cite{Riess04} SNIa data
for a 9 parameter fit with evolving EOS. The dashed lines
correspond to the 68\%$(1\sigma)$ and 90\%$(1.64\sigma)$
confidence intervals.} \label{fig:CLw0wa}
    \end{figure}

\hspace{5mm}
The results of this frequentist combination of CMB and SNIa data are
summarised in Tab.~\ref{tab:DataResults}. When the equation of state
is considered constant, we obtain $w_0 = -0.92_{-0.13}^{+0.10}$
(1-$\sigma$) and the shape of the CL is relatively symmetrical around
the value of $w_0$ obtained at the $\chi^2$ minimum.  When a $z$
dependence is added to the equation of state, the CL is still
symmetrical with $w_0 = -1.09_{-0.15}^{+0.13}$ but $w_a$ becomes
asymmetrical with a long tail for smaller values of $w_a$, as can be
seen in Fig.~\ref{fig:CLw0wa}.  The 1-D CL for $w_a$ gives the
resulting CL at 68\%$(1\sigma)$ and 95\%$(2\sigma)$: $w_a =
0.82_{-0.26}^{+0.21}\,_{-0.80}^{+0.42}$.

\begin{table}[htb]
\begin{center}
\caption{ Results of the frequentist fit to WMAP and Riess et al.
(2004) SNIa data. For the 8 parameter fit with a constant EOS, the
first column gives the value of the variable at the $\chi^2$
minimum, with the confidence interval at 68\% (1 $\sigma$), the
second column shows the $1\sigma$ error computed with the Fisher
matrix techniques. The third and fourth columns present the same
information for the 9 parameter fit with a $z$ dependent EOS. The
$1\sigma$ errors are symmetrical
 for all the variables except for $w_a$. Its error goes from
$_{-0.26}^{+0.21}$ for CL at 68\%  to $_{-0.80}^{+0.42}$ for CL at
95\% (see text).} \label{tab:DataResults}
\begin{tabular}{lcc|cc}
\hline \hline & \multicolumn{2}{c|}{\bf constant EOS} &
\multicolumn{2}{c}{\bf $z$ dependent EOS}\\

 & fit & $\sigma_{Fisher}$& fit & $\sigma_{Fisher}$\\
\hline ${\mathbf \Omega_{b} }$ &  $0.049_{-0.003}^{+0.005}$ & $\pm
0.003$
&  $0.055_{-0.003}^{+0.003}$ & $\pm 0.003$ \\
${\mathbf \Omega_{m} }$ & $0.29_{-0.04}^{+0.05}$ & $\pm 0.04$
 & $0.33_{-0.04}^{+0.04}$ & $\pm 0.04$ \\
${\mathbf h }$ & $0.69_{-0.02}^{+0.03}$ & $\pm 0.03$
& $0.69_{-0.02}^{+0.03}$ & $\pm 0.03$ \\
${\mathbf n_S }$ & $0.97_{-0.03}^{+0.03}$ & $\pm 0.03$
& $0.97_{-0.03}^{+0.03}$ & $\pm 0.03$ \\
${\mathbf \tau }$ & $0.13_{-0.04}^{+0.04}$ & $\pm 0.04$
& $0.14_{-0.04}^{+0.04}$ & $\pm 0.04$\\
${\mathbf w_{0} }$ & $-0.92_{-0.13}^{+0.10}$ & $\pm 0.11$
& $-1.09_{-0.15}^{+0.13}$ & $\pm 0.14$ \\
${\mathbf w_{a} }$ & - & -
&$0.82_{-0.26}^{+0.21}$ & $\pm 0.25$ \\
${\mathbf A }$ & $0.79_{-0.07}^{+0.08}$ & $\pm 0.10$ &
$0.80_{-0.07}^{+0.08}$ & $\pm 0.10$ \\
${\mathbf M_{s0} }$ & $15.94_{-0.03}^{+0.03}$ & $\pm 0.03$
&  $15.95_{-0.03}^{+0.03}$ & $\pm 0.03$\\
 \hline
\end{tabular}
\end{center}
\end{table}

Tab.~\ref{tab:DataResults} compares the $1\sigma$ errors obtained with
the frequentist method and the errors predicted with the Fisher matrix
techniques. The agreement is good, and in the remaining part of this
paper, for the combination of expectations from future surveys, we
will use the Fisher matrix approach.

\begin{figure}
   \centering
  \includegraphics[width=8cm]{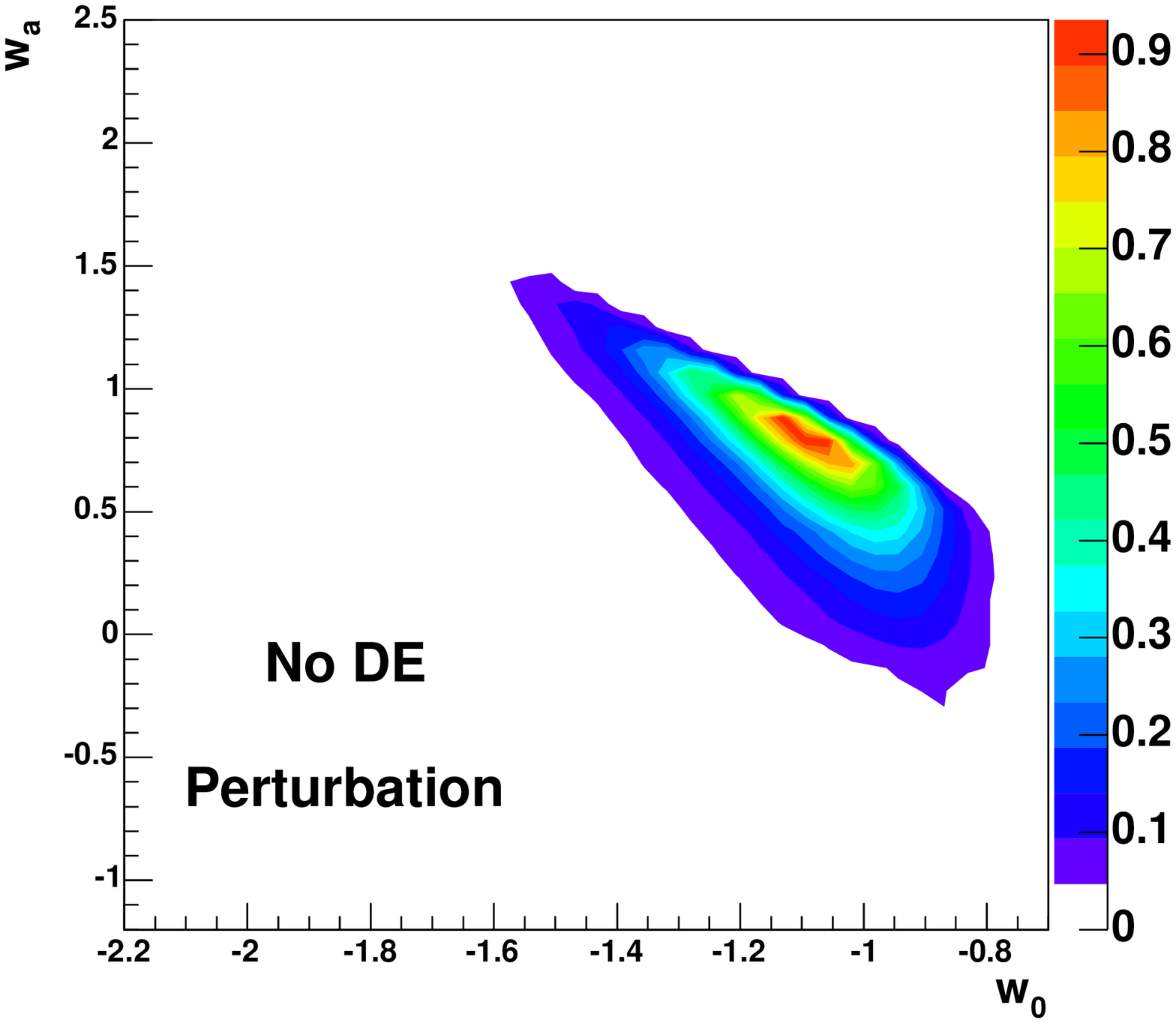}
 \includegraphics[width=8cm]{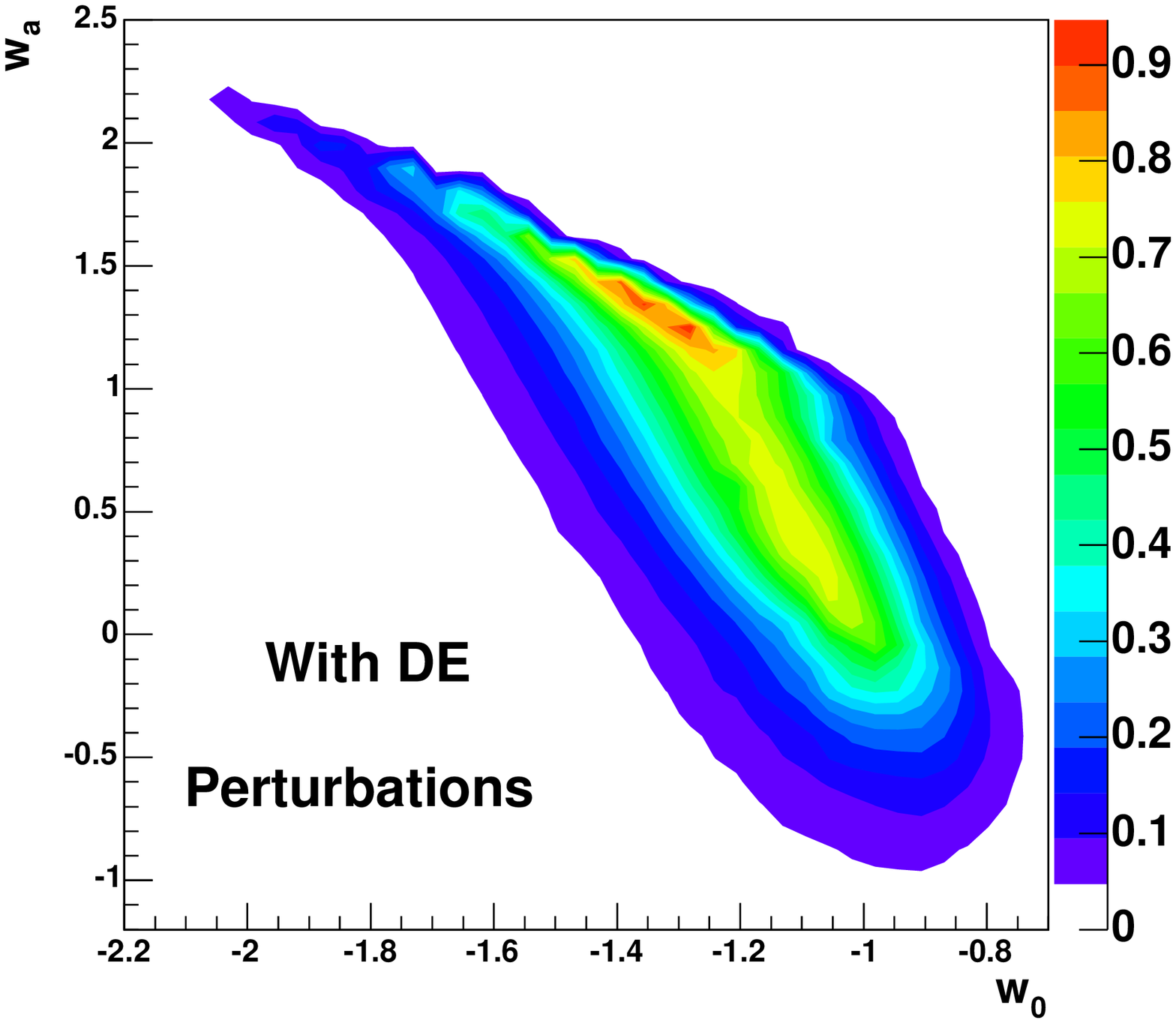}
   \caption{\footnotesize Confidence level contour plots with WMAP
and Riess et al.~\cite{Riess04} SNIa data, for the 9 parameter fit
with a $z$ dependent EOS
 in the plane ($w_0, w_a$). The plot on the left hand side corresponds to the
 case when we introduce no dark energy perturbation. For the plot on the right hand side,
we introduce dark energy perturbations only when $ w > -1$.}
\label{fig:Contourw0wa}
    \end{figure}

However Upadhye et al. (\cite{Upadhye}) noticed that the high
redshift limit of the parametrization of the EOS plays an
important role when we consider CMB data which impose
$w(z\to\infty)<0$.  With our choice of parametrization (see
definition in Eq.~\ref{eq:w0wa}), we get the condition $w_0 + w_a
<0$. When a fit solution is found close to this boundary
condition, as is the case with the current data, the CL
distributions are asymmetric, giving asymmetrical errors. The
Fisher matrix method is not able to represent complicated 2-D CL
shapes, as those shown in Fig.~\ref{fig:Contourw0wa}.  For
example, the error on $w_a$ increases when the $(w_0,w_a)$
solution moves away from the 'unphysical' region $w_0+w_a > 0$. To
avoid this limitation, we will thus use fiducial values of $w_a$
closer to zero for the prospective studies
with future surveys.\\


It is worth noting that the solution found by the fit corresponds to a
value of $w$ slightly smaller than -1 for $z=0$, and a value of $w$
slightly larger than -1 for high $z$. The errors are such that the
value of $w$ is compatible with -1. However, this technically means
that the Universe crosses the phantom line in its evolution.  This
region ($w < -1$) cannot be reached by the fit, if dark energy
perturbations are computed in the CMBEASY version we use.  To obtain a
solution and compare with other published results, we therefore probed two
different conditions, both illustrated in Fig.~\ref{fig:Contourw0wa}.\\

First, we removed altogether the perturbations for the dark
energy, which gives the results presented above.  This allows a
comparison with Seljak et al. (\cite{Seljak}), who have likely
removed dark energy perturbations. Their central value corresponds
to $w_0 = -0.98_{-0.37}^{+0.38}$ and $w_a = -0.05_{-1.13}^{+1.92}
$ at 95\%$(2\sigma)$.
It is closer to $w=-1$ than our result and gives errors for $w_a$
larger than the ones we get.  The comparison is however not exact,
since Seljak et al. use a bayesian approach for the fits, and give
results for an evolving equation of state, only for the total
combination of the WMAP and SNIa data with other SDSS probes
(galaxies clustering, bias, and Lyman $\alpha$ forest).\\

We also performed the fits, including dark energy perturbations,
only when $w > -1$ (which is the default implementation in
CMBFAST). Caldwell \& Doran (\cite{DoranCaldwell}) have argued
convincingly that crossing the cosmological constant boundary
leaves no distinct imprint, i.e., the contributions of $w < -1$
are negligible, because $w < -1$ dominates only at late times and
dark energy does not generally give strong gravitational
clustering. Our analysis, including dark energy perturbations only
when $w > -1$, gives a minimum (cf. right hand side plot in
Fig.~\ref{fig:Contourw0wa}) for $w_0 = -1.32_{-0.19}^{+0.15}$ and
$w_a = 1.2_{-0.8}^{+0.5} $ at $1 \sigma$. This is some $2 \sigma$
away from the no perturbation case. We remark that these values
are very close to those obtained by Upadhye et
al.(\cite{Upadhye}), who use a procedure similar to ours, without
any marginalisation on parameters, a weak constraint $w_0+w_a \le
0$ inside their fit.  Their result, $w_0 = -1.3_{-0.39}^{+0.34}$
and $w_a = 1.25_{-2.17}^{+0.40} $ at 95\%$(2\sigma)$, has almost
the same central value as our fit, when we switch on the dark
energy perturbation for $w > -1$.  The errors we get are also
compatible, and are much larger than in the no perturbation case.

The importance and impact of introducing dark energy perturbations has
been discussed by Weller \& Lewis (\cite{Weller}).  Their combined
WMAP and SNIa analysis with a constant sound speed also gives a more
negative value of $w$, when a redshift dependence is taken into
account.  Although Rapetti et al. (\cite{Rapetti}) observe a reduced
effect when they add cluster data, they still indicate a similar
trend. Finally, when dark energy perturbations are included, we
observe that the minimisation is more difficult and correlations
between parameters increase.\\

We conclude that our results are compatible with other published
papers using various combinations of cosmological probes.  There is a
good agreement of all analysis when $w_0$ is constant, showing that
data agree well with the $\Lambda CDM$ model.  However, large
uncertainties remain for the location of the minimum in the
($w_0,w_a$) plane, when a redshift variation is allowed.  We emphasise
that this is not due to the statistical method but to internal
assumptions.  Upadhye et al.(\cite{Upadhye}) mention the sensitivity
to the choice of parametrization. We show that the introduction of
dark energy perturbations for $w > -1$, can change the minimum by
nearly 2$\sigma$ and that the minimum is not well established as
correlations between parameters increase, and errors, in this zone of
parameter space are very large.

For the sake of simplicity, we decided to present, in the rest of
this paper, a prospective study without dark energy perturbations,
using a Fisher matrix technique.


\section{Combination of future surveys }
\label{sec:MidTerm}
\hspace{5mm}
In this section, we study the sensitivity of the combination
of future CMB, SNIa and weak lensing surveys for dark energy evolution.
We expect new measurements from the CHTLS surveys in SNIa and weak
lensing in the next few years, which can be combined with the
first-year WMAP together with the expected CMB data from the Olimpo
CMB balloon experiment. These are what we call 'mid term' surveys.

The combined mid term results will be compared to the 'long term' expectations
from the next generation of observations in space which are under
preparation, i.e., the Planck Surveyor mission for CMB, expected in
2007, and the SNAP/JDEM mission, a large imaging survey, expected for
2014, which includes both SNIa and weak lensing surveys.

\subsection{Mid term surveys}
\hspace{5mm}
The different assumptions we use for the mid term simulations are as
follows, and are summarised in Tab.~\ref{tab:Surveysim}.
\paragraph*{\bf CMB:\,}
We add to the WMAP data, some simulated CMB expectations from the
Olimpo balloon experiment (Masi et al.~\cite{Olimpo}), equipped
with a 2.6~m telescope and 4 bolometers arrays for frequency bands
centered at 143, 220, 410 and 540~GHz. This experiment will also
allow us to observe the first "large" survey of  galaxies  cluster
through the SZ  effect. For this paper, we will limit our study to
CMB anisotropy aspects.

For a nominal 10 days flight with an angular resolution
$\theta_{fwhm}= 4'$ and with $f_{sky}\simeq 1 \%, $ the expected
sensitivity per pixel is $s=3.4\times10^{-6}$. We use
Eq.~\ref{eq:cmberror} to estimate the statistical error
$\sigma_{C_{\ell}}$ on the angular power spectrum.

\paragraph*{\bf SNIa:\,}
We simulate future SNIa measurements derived from the large SNLS
(\cite{SNLS}) ground based survey within the CFHTLS (\cite{cfht}).
This survey has started in 2003 and expects to collect a sample of
700 identified SNIa in the redshift range $0.3<z<1$, after 5 years
of observations.  We simulate the sample, as explained in Virey et
al. (\cite{biasSN}) with the number of SNIa shown in
Tab.~\ref{tab:stat}, in agreement with the expected SNIa rates
from SNLS. We assume a magnitude dispersion of 0.15 for each
supernova, constant in redshift after all corrections. This
uncertainty corresponds to the most favourable case in which
experimental systematic errors are not considered.

A set of 200 very well calibrated  SNIa at redshift $< 0.1$ should
be measured by the SN factory (Wood-Vasey et al.~\cite{SNfactory}) project.
This sample is needed to normalise the Hubble diagram and will be
called the 'nearby' sample.

\begin{table}[htb]
\begin{center}
\caption{Number of simulated SNIa by bins of 0.1 in redshift for SNLS+HST and SNAP respectively.}
\begin{tabular}{l|cccccccccccccccc}
  \hline \hline
  {\bf z} & 0.2 & 0.3 & 0.4 & 0.5 & 0.6 & 0.7 & 0.8 & 0.9 & 1.0 & 1.1 & 1.2 & 1.3 & 1.4 & 1.5 & 1.6 & 1.7 \\ \hline
{\bf SNLS + HST} & -   & 44  & 56  & 80  &  96 & 100 & 104 & 108 & 10  & 14   & 7   & 12   & 5   & 2   & 3 & 1 \\
{\bf SNAP} & 35  & 64  &  95 & 124 & 150 & 171 & 183 & 179 & 170 & 155 & 142 & 130 & 119 &  107& 94 & 80 \\
  \hline
\end{tabular}
\label{tab:stat}
\end{center}
\end{table}

Finally, to be as complete as possible, we simulate a set of 54 SNIa, expected
from HST programs, with a magnitude dispersion of 0.17 for each supernova,
at redshifts between 1 and 1.7.  Tab.~\ref{tab:Surveysim} summarises the
simulation parameters.

\paragraph*{\bf Weak lensing:\, }
The coherent distortions that lensing induces on the shape of
background galaxies have now been firmly measured from the ground
and from space. The amplitude and angular dependence of this
`cosmic shear' signal can be used to set strong constraints on
cosmological parameters.

Earlier studies of the constraints on dark energy from generic weak
lensing surveys can be found in Hu \& Tegmark (\cite{HuTeg99}),
Huterer (\cite{Hut01}), Hu (\cite{Hu02}).
More recently, predictions for the constraints on an evolving $w(a)$
were studied by several authors (e.g., Benabed \& van Waerbeke
\cite{Benabed}, Lewis \& Bridle \cite{LewisBridle}).  We expect, in
the near future, new cosmic shear results from the CFHTLS wide survey
(CFHTLS ~\cite{cfht}).

In this paper, we will consider measurements of the lensing power
spectrum $C_{\ell}$ with galaxies in two redshift bins. We will only
consider modes between $\ell=10$ and $20000$, thus avoiding small
scales where instrumental systematics and theoretical uncertainties
are more important.
\label{lrange}

For the CFHTLS survey, we assume a sky coverage of
$170^{\circ 2}$.  The rms shear error per galaxy is taken as
$\sigma_{\gamma}=0.35$ and the surface density of usable galaxies
as $20\,\rm amin^{-2}$ which is divided evenly into to redshift bins
with median redshifts $z_m=0.72$ and $1.08$. The redshift distribution
of the galaxies in each redshift bin is taken to be as in Bacon et
al. (\cite{bac00}) with the above median redshifts (cf
Tab.~\ref{tab:Surveysim} for a summary of the survey parameters).
We use Eq.~\ref{eq:wlerror} to estimate the statistical error $\sigma_{C_{\ell}}$.

\begin{table}[htb]
\begin{center}
\caption{
Simulation inputs for CMB, SNIa and
Weak Lensing observations } \label{tab:Surveysim}
\begin{tabular}{l|c|cccc}
 \hline \hline
 \multicolumn{6}{c}{\bf CMB surveys} \\
\hline
& & $\mathbf{ f_{sky}}$  & $ \mathbf{ f(GHz)}$ & $\mathbf{ \theta_{fwhm} (\prime)}$ & $\mathbf{ s(10^{-6})}$ \\
{\bf Current } & WMAP (Spergel et al.(\cite{WMAPSpergel})) &  full sky  & 23/33/41/61/94  & ~13 &  -  \\
{\bf Data } &  &   &  &   &    \\
&  &   &    &  &   \\
{\bf Mid term } & Olimpo &  0.01 &  143/220/410/540 & 4 & 3.4   \\
{\bf Data } & + WMAP &   &  &   &    \\
&  &   &    &  &   \\
{\bf Long term} & &  full Sky & 100 & 9.2 & 2.0  \\
{\bf data } & Planck  &   & 143 & 7.1 & 2.2 \\
{\bf     } &  &   & 217 & 5.0 & 4.8   \\

\hline
\multicolumn{6}{c}{\bf SN surveys} \\
\hline
& & {\bf SN \# } & {\bf Redshift range} & {\bf Statistical error} &  {\bf Systematic error} \\
{\bf Current } & Riess et al.~(\cite{Riess04}) + HST &  157 & $z < 1.7$ & $\sim 0.25$  &  - \\
{\bf Data } &   &    &   &  &  \\
 &  &   &    &  &   \\
{\bf Mid term } & SNfactory &  200 & $z < 0.1$ & 0.15  &  - \\
{\bf Data } & SNLS &   700 &   $ 0.3 < z < 1 $ & 0.15 & - \\
& HST  &  54 &   $ 1 < z $ & 0.17  & \\
 &  &   &    &  &   \\
{\bf Long term }  & SNfactory    & 300  & $z < 0.1$ & 0.15 &  \\
{\bf Data } & SNAP &  2000 & $ 0.1 < z < 1.7 $  & 0.15 &  0.02 \\

\hline
 \multicolumn{6}{c}{\bf WL surveys} \\
 \hline
& & $\mathbf z_{m} $ {\bf (2 bins)}  & $\mathbf{ A (\deg^2)}$ &{\bf total} $\mathbf{ n_g (\rm{\bf amin}^{-2})}$ & $\mathbf{ \sigma_{\gamma}}$  \\

{\bf Mid term} & CFHTLS &  0.72, 1.08 &  170  & $ 20 $ & 0.35    \\
{\bf Data} &  &   &    &  &   \\

 &  &   &    &  &   \\
{\bf Long term } & SNAP &  0.95, 1.74 & 1000   &  100 & 0.31  \\
{\bf Data} &    &   &  &  &  \\
 \hline

\end{tabular}
\end{center}
\end{table}

\subsection{Long term survey}
\hspace{5mm}
The future will see larger surveys both from the ground and space.  To
estimate the gain for large ground surveys compared to space, critical
studies taking into account the intrinsic ground limitation (both in
distance and in systematics) should be done, and systematic effects,
not included here, will be the dominant limitation.  In this paper, we
limit ourselves to the future space missions.

We simulate the Planck Surveyor mission using
Eq.~\ref{eq:cmberror} with the performances described in Tauber et
al. (\cite{planck}). Assuming that the other frequency bands will
be used to identify the astrophysical foregrounds, for the CMB
study over the whole sky, we consider only the three frequency
bands (100, 143 and 217 GHz) with respectively  $(\theta_{fwhm}=
9.2',\,\,7.1'\,\,{\rm and}\,\,5.0')$ resolution and
$(s=2.0\,10^{-6},\,\,2.2\,10^{-6}\,\,{\rm and}\,\,4.8\,10^{-6})$
sensitivity per pixel.

We also simulate observations from the future SNAP satellite, a
2\,m telescope which plans to discover around 2000 identified
SNIa, at redshift 0.2$<z<$1.7 with very precise photometry and
spectroscopy. The SNIa distribution, given in Tab.~\ref{tab:stat},
is taken from Kim et al. (\cite{Kim}).  The magnitude dispersion
$\sigma (m)_{disp}$ is assumed to be 0.15, constant and
independent of the redshift, for all SNIa after correction.
Moreover, we introduce an irreducible systematic error $\sigma
(m)_{irr}$ following the prescription of  Kim et al. (\cite{Kim}).
In consequence, the total error on the magnitude $\sigma
(m)_{tot}$ per redshift bin $i$, is defined as: $\sigma
(m)_{tot,i}^2 = \sigma (m)_{disp}^2/N_i + \sigma (m)_{irr}^2$
where $ N_i$ is the number of SNIa in the ith 0.1 redshift bin. In
the case of SNAP, $\sigma (m)_{irr}$ is equal to $0.02$.

The SNAP mission also plans a large cosmic shear survey.  The
possibilities for the measurement of a constant equation of state
parameter $w$ with lensing data were studied by Rhodes et al.
(\cite{Rhodes}), Massey et al. (\cite{massey}), Refregier et al.
(\cite{Refregier}).  We extend here the study in the case of an
evolving equation of state. We use in the simulation the same
assumptions as in Refregier et al. (\cite{Refregier}) with a
measurement of the lensing power spectrum in 2 redshift bins,
except for the survey size, which has increased from $300^{\circ
2}$ to $1000^{\circ 2}$ (Aldering et al. \cite{SNAP}) and for the
more conservative range of multipoles $\ell$ considered (see
\S\ref{lrange}).

The long term survey parameters are summarised in Tab.~\ref{tab:Surveysim}.

\subsection{Results}
\hspace{5mm}
The combination of the three data sets is performed with, and without,
a redshift variation for the equation of state, for both mid term and
long term data sets.

\begin{figure} [htb]
   \centering
  \includegraphics[width=12cm]{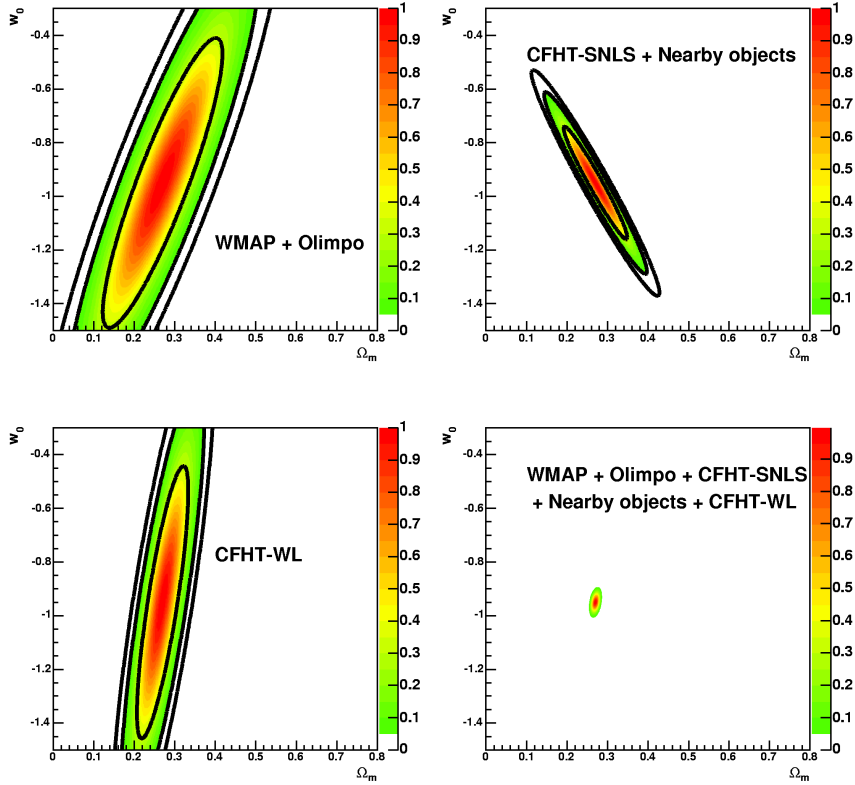}
     \caption{\footnotesize CL contours for mid term CMB (WMAP
+Olimpo), SNIa  and weak lensing data from CFHTLS and the
combination of the three probes for the 8 parameter fit in the plane ($\Omega_m, w_0$)
(see also Tab.~\ref{tab:SummaryResults}). The solid lines represent
68\% (1 $\sigma$), 95\% (2$\sigma$), and 99\% CL contours.}
\label{fig:result3a}
    \end{figure}

The different plots in Fig.~\ref{fig:result3a} show the results for
individual mid term probes and for their combination.  The results are
for a constant $w_0$, plotted as a function of the matter density
$\Omega_m$.  The combined contours are drawn using the full
correlation matrix on the 8 parameters for the different sets of data.

The SNLS survey combined with the nearby sample will improve the
present precision on $w$ by a factor 2. The expected contours from
cosmic shear have the same behaviour as the CMB but provide a slightly
better constraint on $\Omega_m$ and a different correlation with $w$:
CMB and weak lensing data have a positive ($w ,\Omega_m$) correlation
compared to SNIa data, which have a negative correlation. This
explains the impressive gain when the three data sets are combined, as shown in
Tab.~\ref{tab:SummaryResults}. Combining WMAP with Olimpo data, helps
to constrain $w$ through the correlation matrix as Olimpo expects to
have more information for the large $\ell$ of the power spectrum.

\begin{figure} [htb]
   \centering
  \includegraphics[width=12cm]{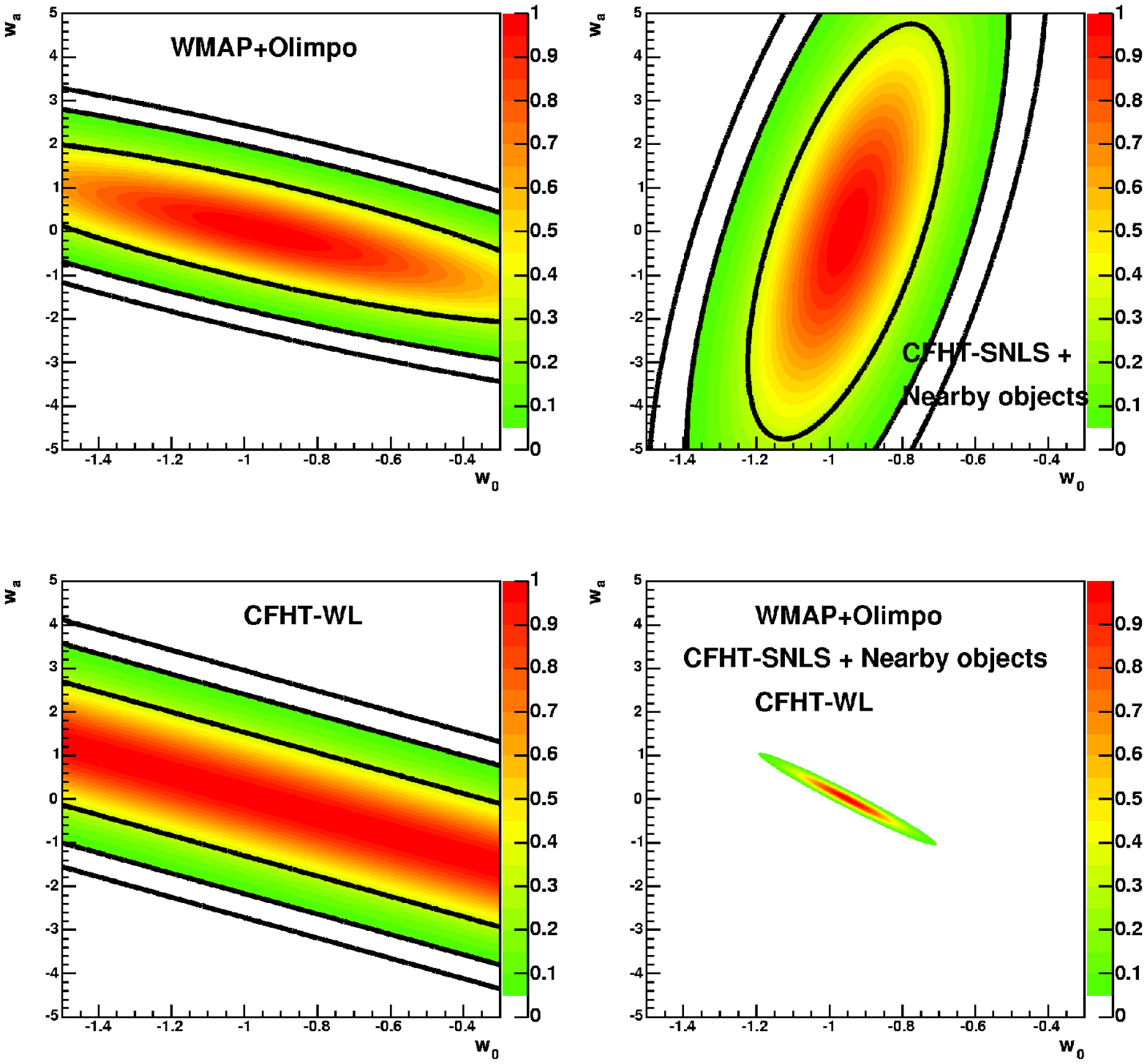}
     \caption{\footnotesize CL contours for mid term CMB (WMAP
+Olimpo), SNIa  and weak lensing data from CFHTLS and the
combination of the three probes for the 9 parameter fit in the plane ($ w_0, w_a$)
 (see also Tab.~\ref{tab:SummaryResults}). The solid lines represent
68\% (1 $\sigma$), 95\% (2$\sigma$), and 99\% CL contours.}
\label{fig:result3b}
    \end{figure}

Fig.~\ref{fig:result3b} gives the expected accuracy of the mid term
surveys on the parameters of an evolving equation of state. The CL
contours plots of $w_a$ versus $w_0$, are obtained with a 9
cosmological parameter fit.  Here also, we observe a good
complementarity: there is little information on the time evolution
from SNIa with no prior, while the large redshift range from CMB data
is adding a strong anti-correlated constraint on $w_a$.

A combined analysis proves far superior to analysis with only SNIa.
In the favourable case, where we add more SNIa from HST survey, we
expect a gain of a factor 2 on the errors, but it is not enough to
lift degeneracies and the expected precision on $w_a$ with these data
will not be sufficient to answer questions on the nature of the dark
energy.

\begin{table}[htb]
\begin{center}
\caption{Expected sensitivity on cosmological parameters for three
scenarii: Current supernova and CMB experiments (WMAP and Riess et
al.\cite{Riess04}), mid term experiments (CFHT-SNLS (supernova
surveys), CFHTLS-WL (weak lensing) and CMB (WMAP+Olimpo)), long
term experiments (CMB (Planck) and SNAP (supernovae and weak
lensing)). For each scenario, the first column gives the $1\sigma$
error computed with the Fisher matrix techniques for the 8 free
parameter configuration and the second columns gives the $1\sigma$
error for the 9 free parameter configuration.
}\label{tab:SummaryResults}
\begin{tabular}{l|cccccc}
\hline \hline \bf{Scenario} & \multicolumn{2}{c|}{\bf Today} &
\multicolumn{2}{c|}{\bf Mid term}
&  \multicolumn{2}{c}{\bf Long Term}\\
\hline
${\mathbf \Omega_{b} }$ & $0.003$ &$0.004$& $0.001$& $0.002$& $0.0008$&  $0.0008$\\
${\mathbf \Omega_{m} }$ & $0.04$ &$0.04$ & $0.01$ & $0.01$& $0.004$& $0.004$ \\
${\mathbf h }$ & $0.03$ &$0.03$ & $0.01$&  $0.01$& $0.006$& $0.006$\\
${\mathbf n_s }$ & $0.03$&$0.03$ & $0.006$& $0.009$& $0.003$& $0.003$\\
${\mathbf \tau }$ &  $0.05$&$0.04$ & $0.01$& $0.01$& $0.01$& $0.01$\\
${\mathbf w_{0} }$ & $0.11$ &$0.22$ & $0.02$& $0.10$& $0.02$& $0.04$\\
${\mathbf w_{a} }$ & $-$& $0.99$ & $-$& $0.43$& $-$ & $0.07$\\
${\mathbf A }$ & $0.10$& $0.10$ & $0.02$& $0.02$& $0.02$&  $0.02$\\
${\mathbf M_{s0} }$ & $0.03$ & $0.03$& $0.01$& $0.01$& $0.01$ & $0.01$\\
 \hline
\end{tabular}
\end{center}
\end{table}

\begin{figure}[htb]
   \centering
    \includegraphics[width=8cm]{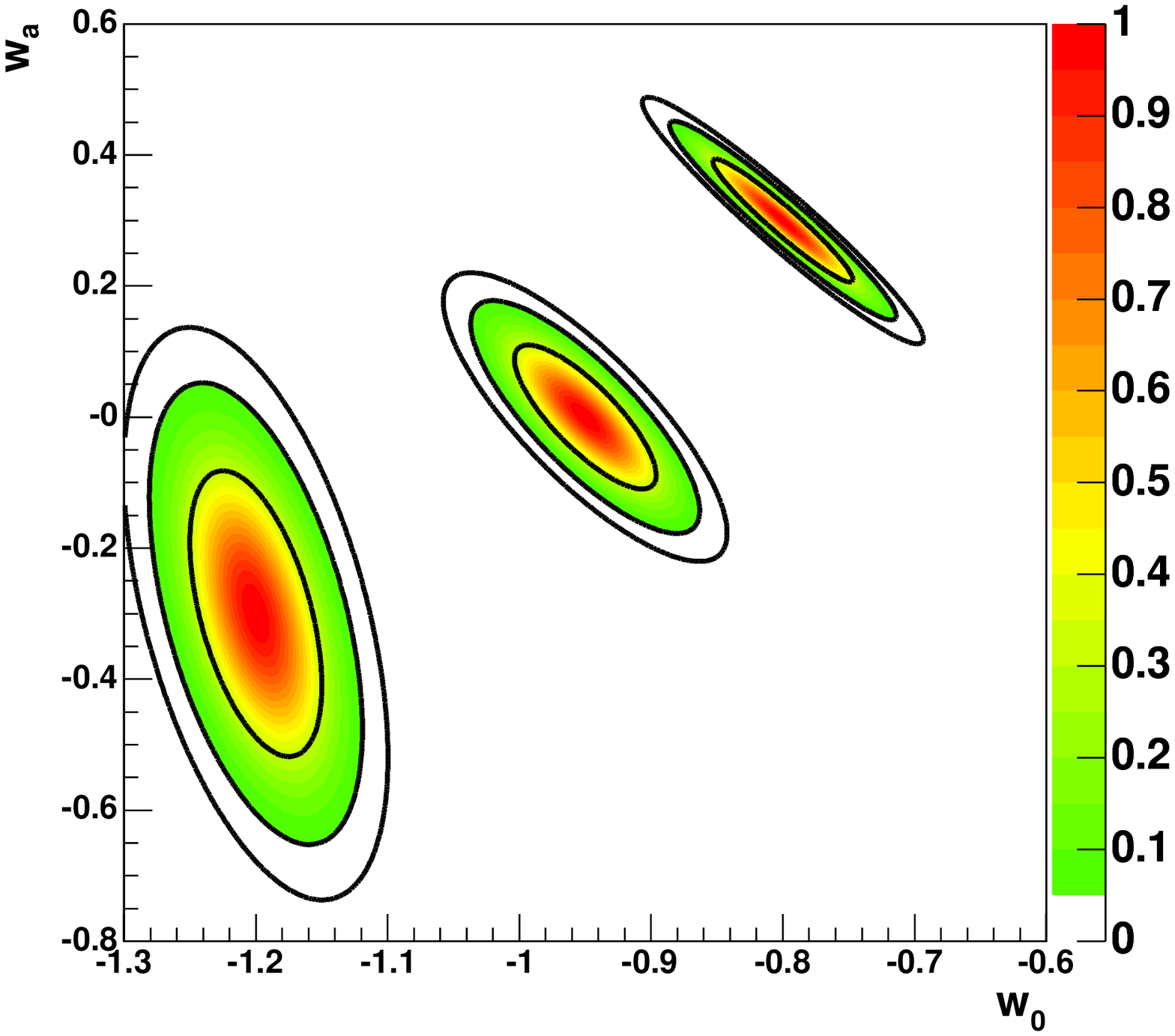}
    \includegraphics[width=8cm]{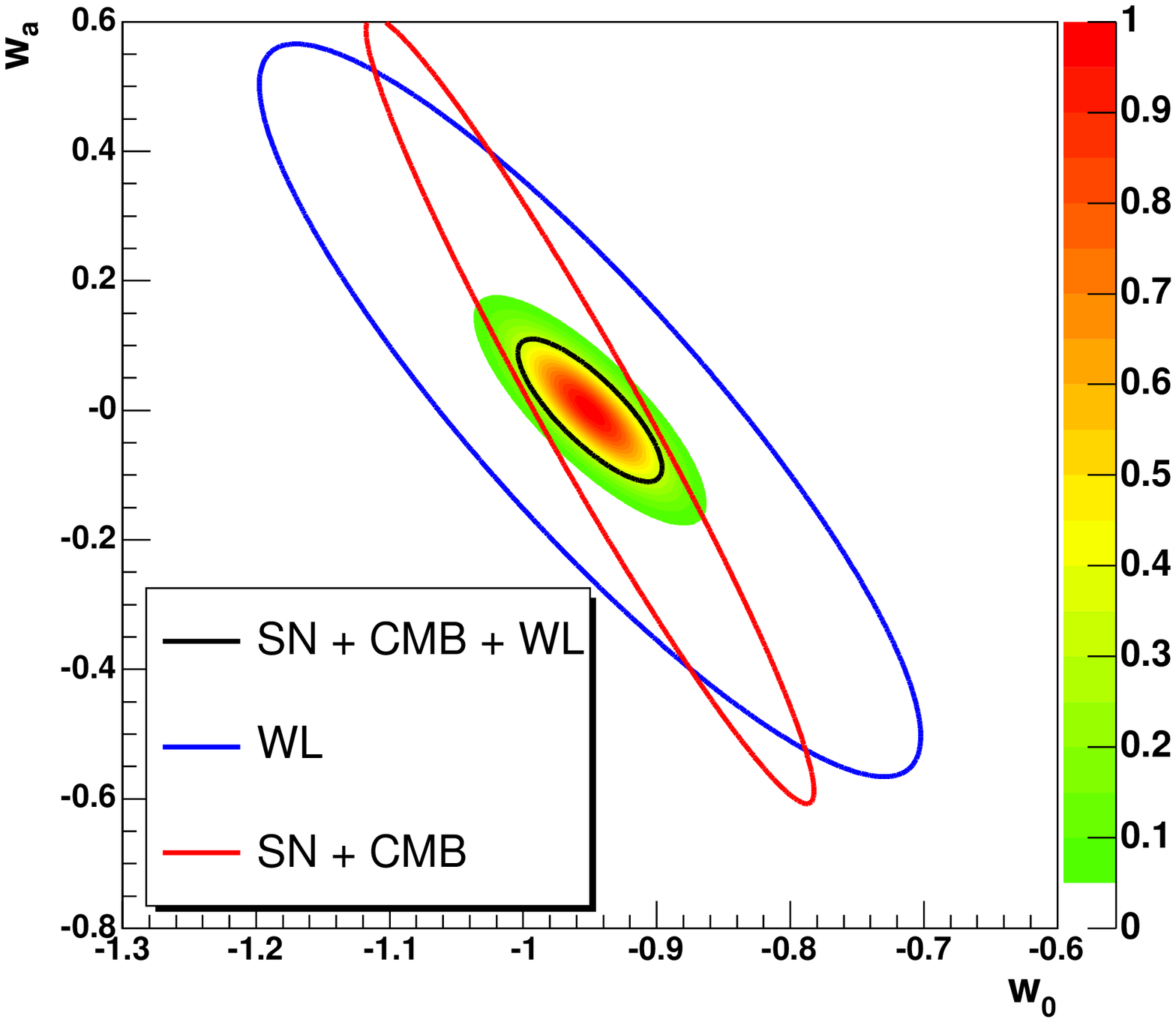}

 \caption{\footnotesize CL contours for future space data from SNAP (SNIa and WL)
and Planck (CMB) for a 9 parameter fit in the plane ($w_0,w_a $).
The left hand side figure shows the combination of  SNAP (SNIa+WL) and CMB
for three different models ($\Lambda$CDM,
SUGRA and Phantom). The solid lines represent
68\% (1 $\sigma$), 95\% (2$\sigma$), and 99\% CL contours. The right hand side
figure shows the CL
for the combined three "long term" probes. The solid lines are the 1$\sigma$ contours for different combinations:
WL alone, combined SNIa and CMB, and the three combined probes.}
\label{fig:result5}
  \end{figure}

The simulated future space missions show an improved sensitivity
to the time evolution of the equation of state. The accuracy on
$w_a$ for the different combinations are summarised in
Tab.~\ref{tab:SummaryResults}.  There is again a large improvement
from the combination of the three data sets.  The precision, for
the long term surveys, will be sufficient to discriminate between
the different models we have chosen, as shown in the left hand
side plot of Fig.~\ref{fig:result5} and in
Tab.~\ref{tab:ModelResults}, while it is not the case for the mid
term surveys. This figure illustrates, moreover, that the errors
on $w_a$ and $w_0$, and the correlation between these two
variables are strongly dependent on the choice of the fiducial
model.

\begin{table}[htb]
\begin{center}
\caption{Expected sensitivity on cosmological parameters for the
long term missions with CMB (Planck) and SNAP (supernova surveys
and weak lensing) for the 9 free parameter configuration.}
\label{tab:ModelResults}
\begin{tabular}{l|ccc}
\hline \hline
\bf{Model} & {\bf $\mathbf \Lambda$CDM} & {\bf SUGRA}&  {\bf Phantom}\\
\hline
${\mathbf \Omega_{b} }$ &  $0.0008$ &  $0.0008$ &  $0.0007$ \\
${\mathbf \Omega_{m} }$ &  $0.004$ &  $0.004$ &  $0.003$ \\
${\mathbf h }$ &  $0.006$ & $0.006$ &  $0.005$ \\
${\mathbf n_S }$ &  $0.003$ & $0.003$ &  $0.003$ \\
${\mathbf \tau }$ &  $0.01$ & $0.01$ &  $0.01$ \\
${\mathbf w_{0} }$ &  $0.04$ &  $0.04$ &  $0.03$ \\
${\mathbf w_{a} }$ & $0.07$ &  $0.06$ &  $0.14$ \\
${\mathbf A }$ &  $0.02$ &  $0.02$ & $0.02$ \\
${\mathbf M_{s0} }$ & $0.015$ & $0.014$ & $0.013$ \\
 \hline
\end{tabular}
\end{center}
\end{table}

More generally, the combination of the probes with the full
correlation matrix allows the extraction of the entire information
available. For instance, the large correlation between $n_S$ and
$w_a$ observed for the weak lensing probe combined with the
precise measurement of $n_s$ given by the CMB, gives a better
sensitivity on $w_a$ than the simple combination of the two $w_a$
values, obtained separately for the CMB and weak lensing.  Such an
effect occurs for several other pairs of cosmological parameters
considered in this study. The plot, in the right hand side of
Fig~\ref{fig:result5}, is an illustration of this effect.  It
shows the combination of the 3 probes in the ($w_0,w_a $) plane.
The $1\sigma$ contour for the combined three probes, is more
constraining than the 2-D combination in the ($w_0,w_a $) plane of
the three probes.

Finally, in the long term scenario, the weak lensing probe
provides a sensitivity on the measurement of $(w_0,w_a)$ comparable
with those of the combined SN and CMB probes, whereas in the mid
term scenario the information brought by weak lensing was marginal.
This large improvement observed in the information provided by the
weak lensing, can be explained by the larger survey size and the
deeper volume probed by SNAP/JDEM, compared to the ground CFHTLS WL
survey. We thus conclude that adding weak lensing information will be an
efficient way to help distinguishing between dark energy models.  If
systematic effects are well controlled, the future dedicated space
missions may achieve a sensitivity of order 0.1 on $w_a$.

The SNAP/JDEM space mission is designed, in principle, to control
its observational systematic effects for SNIa to the $\%$ level,
which is probably impossible to reach for future ground
experiments. In this study, we assign an irreducible systematic
error on SNIa magnitudes of 0.02 and
systematic effects have been neglected for CMB and weak lensing.
This can have serious impacts on the final sensitivity, in particular, on the
relative importance of each probe.

Other probes, whose combined effects we have not presented in this
paper, but intend to do in forthcoming studies, remain therefore
most useful. For example, the recent evidence for baryonic
oscillations (Eisenstein et al. 2005) is a proof that new probes
can be found.  The present constraints that these results provide,
do not improve the combined analysis we present here.  However,
getting similar results from different probes greatly contributes
to the credibility of a result, in particular, when the
systematical effects can be quite different, as is the case for
the different probes we consider. Finally, the joint analysis of
cluster data observed simultaneously with WL, SZ effect and
X-rays, will allow the reduction of the intrinsic systematics of
the WL probe.

\section{Conclusions}
\label{sec:conclusion}

\hspace{5mm}

In this paper, we have presented a statistical method based on a
frequentist approach to combine different cosmological probes. We have
taken into account the full correlations of parameters without any
priors, and without the use of Markov chains.

Using current SNIa and WMAP data, we fit a parametrization of an
evolving equation of state and find results in good agreement with
other studies in the literature.  We confirm that data prefer a
value of $w$ less than -1 but are still in good agreement with the
$\Lambda CDM$ model.  We emphasise the impact of the
implementation of the dark energy perturbations.  This can explain
the discrepancies in the central values found by various authors.
We have performed a complete statistical treatment, evaluated the
errors for existing data and validated that the Fisher matrix
technique is a reliable approach as long as the parameters
$(w_0,w_a)$ are in the `physical' region imposed by CMB boundary
condition: $w(z\to\infty)<0$.

We have then used the Fisher approximation to calculate the expected
errors for current surveys on the ground (e.g., CFHTLS) combined with
CMB data, and compared them with the expected improvements from future
space experiments.  We confirm that the complete combination of the
three probes, including weak lensing data, is very powerful for the
extraction of a constant $w$. However, a second generation of
experiments like the Planck and SNAP/JDEM space missions is required,
to access the variation of the equation of state with redshift, at the
0.1 precision level. This level of precision needs to be
confirmed by further studies of systematical effects, especially for
weak lensing.

 \begin{acknowledgements} The authors are most grateful to M. Doran
 for the CMBEASY package, the only code that was not developed by this
 collaboration, and for his readiness to answer all questions. They
 wish to thank  A. Amara, J. Berg\'{e}, A. Bonissent, D. Fouchez, F. Henry-Couannier,
 S. Basa, J.-M. Deharveng, J.-P. Kneib,  R. Malina,
 C. Marinoni, A. Mazure, J. Rich, and P. Taxil for their
 contributions to stimulating discussions.
\end{acknowledgements}


\end{document}